# Bremsstrahlung of 5–25 keV electrons incident on $MoSi_2$, $TiB_2$ and $ZrB_2$ thick solid conductive compounds

Heng Zhang, Zhu An[*], Jingjun Zhu[*], Hong Huang

Key Laboratory of Radiation Physics and Technology of Ministry of Education, Institute of Nuclear Science and Technology, Sichuan University, Chengdu 610064, China

[*] Corresponding authors.
E-mail addresses: anzhu@scu.edu.cn (Z. An), zhujingjun@scu.edu.cn (J. Zhu).

**Abstract**

Absolute measurements were conducted to study the bremsstrahlung emission from ~5-25 keV electrons incident on three thick solid conductive compounds of $MoSi_2$, $TiB_2$ and $ZrB_2$. The additivity approximation was applied in the Monte Carlo PENELOPE simulations for compounds and mixtures. The results showed that in general the experimental bremsstrahlung spectra were in good agreement with the Monte Carlo simulation results, suggesting the feasibility of the additivity approximation in Monte Carlo simulations for the studied cases even in the absolute measurements and that the significant differences between experiments and Monte Carlo simulations near the Duane-Hunt limit for insulating targets in previous studies do not appear in the present studies.

# 1. Introduction

When a charged particle passes by an atomic nucleus, its speed is changed due to the Coulomb interaction and the electromagnetic radiation is emitted. This radiation is called bremsstrahlung (BS). BS is an important part of the interaction between charged particles and matter. The in-depth study and accurate understanding of BS are of great significance to the researches in many fields, such as materials science, astrophysics and radiology, etc. [1]. While there have been many investigations of bremsstrahlung emission resulting from electron interactions with single elemental targets, including metallic targets and rare gas targets [2-9], there have been much fewer experiments performed for studying bremsstrahlung emitted by electrons interacting with molecular (or compound) targets [10-15]. It was observed that the experimental bremsstrahlung spectra are in very good agreement with the Monte Carlo (MC) simulation results for single elemental metallic targets [16-18], including Ti, Mo and Zr elements, and the adjacent elements for B and Si, i.e., C and Al elements.

For compounds and mixtures, in 1905, Bragg and Kleeman first applied the Bragg additivity approximation [19], which states that the stopping power of a compound can be estimated by the linear addition of the stopping powers of its individual elements. The accuracy of the Bragg additivity approximation is limited by its neglect of the changes of orbital structure and excitation structure caused by the differences of bonding in single elemental materials and compounds. The Bragg additivity approximation assumes that the average ionization energy of each atom remains constant regardless of its chemical state in the compound molecule. Bragg additivity approximation generally performs well for calculating the stopping power of medium and high energy particles. However, significant deviations from experimental data still have been observed for some compounds and mixtures in a wide incident electron energy region [20-24].

In MC simulations, besides for the stopping powers, the additivity approximation is also used to treat all cross sections involved for compounds and mixtures, including the bremsstrahlung differential cross sections, for example, in MC PENELOPE code [25]. Previous studies on BS radiation from molecular targets mainly involved gaseous targets using relative measurements in the incident electron energies of 3.5-15 keV, and it was shown that the experimental results were in good agreement with the theoretical results using the additivity approximation [10-15]. However, the additivity approximation may also be affected by the phase state [26]. Acosta et al [27] measured the thick-target bremsstralung spectra for a Fe-Cr-Ni alloy and a ZnS at incident electron energies of 20 and 30 keV and found that the experimental spectra were in good agreement with the PENELOPE's predictions, where the detector efficiency was determined based on the geometrical parameters given by the detector's manufacturer and the reliability of Monte Carlo simulations. Most recently, Adamson et al. [28-30] investigated the bremsstrahlung emission from 5 keV electrons incident on insulating solid compound targets, i.e., BeO, NaCl, $Al_2O_3$, MgO and $SrTiO_3$, using relative measurements. They compared the experimental results with MC PENELOPE [25] simulations and found that they agreed well in low and medium X-ray energy regions but experimental results are significantly larger than the simulation results near the Duane-Hunt limit. It was inferred that this difference was likely due to the charging effects in the insulating targets or the additivity approximation [29,30]. Most of these previous studies for compounds [10-15,27-30] were conducted in the ways of relative measurements, except the work of Acosta et al [27].

For further distinguishing whether the charging effects in the insulating targets or the additivity approximation are responsible for the difference between the previous experiments and the MC simulations [28-30], this work will investigate three single-phase solid conductive compounds, i.e., molybdenum disilicide ($MoSi_2$), titanium diboride ($TiB_2$) and zirconium diboride ($ZrB_2$). The three solid compounds have very good electrical conductivity and a wide range of applications in industries [31-36].

The computer code system PENELOPE [25] performs Monte Carlo simulations of coupled electron-photon transport in arbitrary materials for a wide energy range from 50 eV to 1 GeV. In PENELOPE, the additivity approximation is used for compounds and mixtures, including the stopping powers and the bremsstrahlung differential cross sections, etc. The difference between the experimental and simulation spectra for compounds and mixtures can be used to validate the feasibility of the additivity approximation used in MC simulations.

In this work, the *absolute* bremsstrahlung spectra of thick $MoSi_2$, $TiB_2$ and $ZrB_2$ samples by electron impact are measured and compared with the simulation spectra of the MC PENELOPE code. The purpose of this work is to obtain more absolute experimental data and investigate the additivity approximation while avoiding the interferences of charging effects on insulating compounds, i.e., to further test the ability of the MC simulations to describe the bremsstrahlung spectra for compounds and mixtures obtained in absolute measurements.

In this paper, Section 2 introduces the experiments and Monte Carlo simulations; Section 3 compares the experimental results with the Monte Carlo simulations and discusses the results; the conclusions are summarized in Section 4.

## 2. Experiments and Monte Carlo simulations

### 2.1 Experimental Setup

The experimental setup was similar to that used in our previous studies [1,17]. A scanning electron microscope (SEM) (KYKY-2800B, KYKY technology Co., Ltd., China) produced a focused electron beam with an energy range of ~5-25 keV. The incident electron energies were determined by using linear extrapolation to find the spectral intercept to zero intensity, i.e., the Duane-Hunt limit, as done by Aguilar, et al [37]. For the thin-target experiments, more careful method should be taken to determine the incident electron energies [38]. The beam spot size was approximately 2-3 mm in diameter. This beam passed through the top aperture of a Faraday cup and irradiated the compound samples placed at the cup's center. The normal of sample plane was at a 45 ° angle to the incident electron beam. The emitted photons were then collected by a silicon drifted detector (SDD) (XR-100SDD, Amptek, Inc., USA) located outside the side aperture of the Faraday cup. The detector was at a 90 ° angle to the incident electron beam. The Faraday cup and the SDD were in the vacuum chamber of the scanning electron microscope. The pile-up reject circuitry (PUR) and the rise time discriminator (RTD) are not active. A bias voltage of -100 V was applied to the top and side apertures of the Faraday cup to prevent the escape of secondary electrons. A magnetic field (~800 G) provided by a pair of permanent magnets was set in front of the SDD to deflect the backscattered electrons escaping through the side aperture. Incident charges was measured using an Ortec 439 digital current integrator and an Ortec 996 counter. The charge measurement system was verified using a Keithley 6430 Sub-Femtoamp Remote Source Meter and an uncertainty of less than 1% was found. The compound samples were obtained from Beijing Goodwill Metal Co., Ltd. and were fabricated by hot pressing of powder. The samples have a diameter of 31 mm and a thickness of 1 mm. The continuous slowing-down approximation (CSDA) ranges for the three samples at the 25 keV were calculated using the code SBETHE [39], and they were less than 4.1 μm, and much smaller than the thickness of the targets.

The SDD used in the experiment has a resolution of 135 eV full-width at half maximum (FWHM) at 5.89 keV of Mn Kα line at 11.2 μs peaking time with an effective area of 25 mm$^2$ and a Si sensitive layer thickness of 500 μm. It contains a multilayer collimator (ML) composed of a tungsten (W) base metal of 100 μm, a chromium (Cr) layer of 35 μm, a titanium (Ti) layer of 15 μm and an aluminum (Al) layer of 75 μm. It also has an ultra-thin C2 detector window composed of 40 nm of silicon nitride, 30 nm of grounded aluminum coating and 14 μm of silicon mesh, and the ratio of the opening area to the total area of the C2 window is 75%. The detector's efficiency calibration was carried out in our previous studies [17-18,40]. In this work, the efficiency curve of the detector was re-examined. The standard point sources were used, i.e., a $^{137}$Cs source produced by the Physikalisch-Technische Bundesanstalt, Germany (PTB), two $^{241}$Am sources produced by Eckert & Ziegler, USA and National Institute of Metrology, China, respectively. The half-lives, X-ray energies and intensity emission rates of the standard sources were adopted from the recommended values of the International Atomic Energy Agency (IAEA) [41]. The average efficiency calibration curve was used as the detector's efficiency values in this work, which have a difference of ~3% compared with the previous efficiency values [17,18] and this difference was used as the uncertainty of efficiency calibration curve.

## 2.2 Sample characterization

Rutherford backscattering spectrometry (RBS) is one of the most widely used ion beam analysis methods. It is a powerful tool for surface elemental analysis of samples, with the advantages of being rapid, non-destructive, multi-elemental and accurate (typically ~3%) [42]. RBS is typically used to detect heavy elements on light elemental matrix, while it is insensitive to light elements on heavy elemental matrix [43]. Elastic backscattering spectrometry (EBS) is a similar technique to RBS, but it uses proton beams instead of helium beams. Proton beams have larger cross sections for the light elements than helium beams, which make EBS more sensitive to light elements [44]. Due to the existence of B element in the heavy elemental matrices for the studied samples, EBS was used to verify the elemental compositions of the three compound samples.

Particle-induced X-ray emission (PIXE) is another widely used ion beam analysis method which utilizes characteristic X-rays induced via ion beam interactions with the target atoms. PIXE is a very sensitive analytical technique because of high X-ray yields and relatively low bremsstrahlung backgrounds, with a sensitivity of 1-100 ppm, depending on atomic number and matrix. In this study, PIXE was also used to qualitatively identify the impurities in the three compound samples as a complementary analysis to the EBS.

EBS and PIXE experiments were simultaneously conducted on the 3 MV tandetron accelerator at the Institute of Nuclear Science and Technology of Sichuan University [45]. 2 MeV proton beams were used and impacted vertically on the samples. In EBS, an ion-implanted-silicon detector (ORTEC U-012-050-100) was placed at a position of 165° relative to the incident proton beam. The current was set to 1 to 2 nA to reduce pile-up effects. The beam spot diameter was approximately 1 mm. The vacuum was $1.6 \times 10^{-7}$ mbar. The dead time was about 4%. At the same time, in PIXE, a low-energy germanium (LEGe) detector (CANBERRA GL0055) with 25 μm Be window and 5 mm sensitive layer was placed at 135°. The typical spectrum of PIXE for $MoSi_2$ is shown in Figure 1. Moreover, the wavelength-dispersive electron probe micro-analysis (WD-EPMA) was also used to analyze the samples, mainly for lighter elements (such as O and Al, etc.). The WD-EPMA analyses were conducted on a JEOL JXA-iHP200F field emission electron probe microanalyzer with 15 kV accelerating voltage and 20 nA beam current intensity, equipped with 4 channels wavelength-dispersive spectrometer with different analyzing crystals.

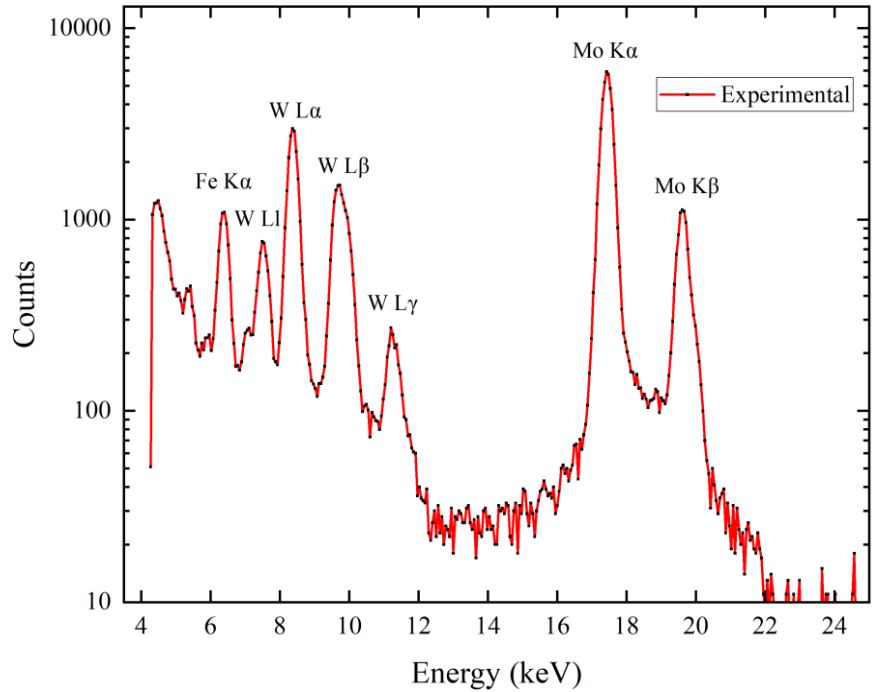

Figure 1. The PIXE experimental spectrum of MoSi$_2$ sample at 135 °using 2 MeV proton impact.

The EBS spectra were analyzed using SIMNRA 7.03 code [46], incorporating multiple scattering, dual scattering, SRIM stopping power data and Chu and Yang energy-loss straggling model. In the EBS analyses, the cross sections of $^{10,11}$B(p,p) were corrected by a factor of 1.19 due to the systematic error in the original measurements [47]. Moreover, the low-energy tails and the slowly-changing shape of high-energy edges of the EBS spectra were produced due to the surface roughness of the samples [48-51], especially for ZrB$_2$ sample at the high-energy edge. The impurities detected by PIXE or WD-EPMA, also simultaneously appeared in our present bremsstrahlung spectra, were incorporated into the EBS simulations. Figures 2-4 show the EBS experimental and simulation spectra. The measured atomic concentrations are as follows: for MoSi$_2$ samples, Mo = 32.43%, Si = 66.49%, W=0.94% and Fe =0.14%; for TiB$_2$ samples, Ti = 31.95%, B = 62.44%, Al = 1.20% and O = 4.41%; for ZrB$_2$ samples, Zr = 32.40%, B= 67.38% and Hf = 0.22%. It is noticed that the concentration ratios of Si/Mo, B/Ti and B/Zr approach to 2 (i.e., 2.05, 1.95 and 2.08, respectively), the differences between these ratios and 2 are less than 5%. The uncertainties of major elemental concentrations are estimated to be 0.3-3.9%, mainly from the uncertainties of EBS cross sections of B and Si [47,52], which will propagate an uncertainty of ~3% to the measured bremsstrahlung spectra.

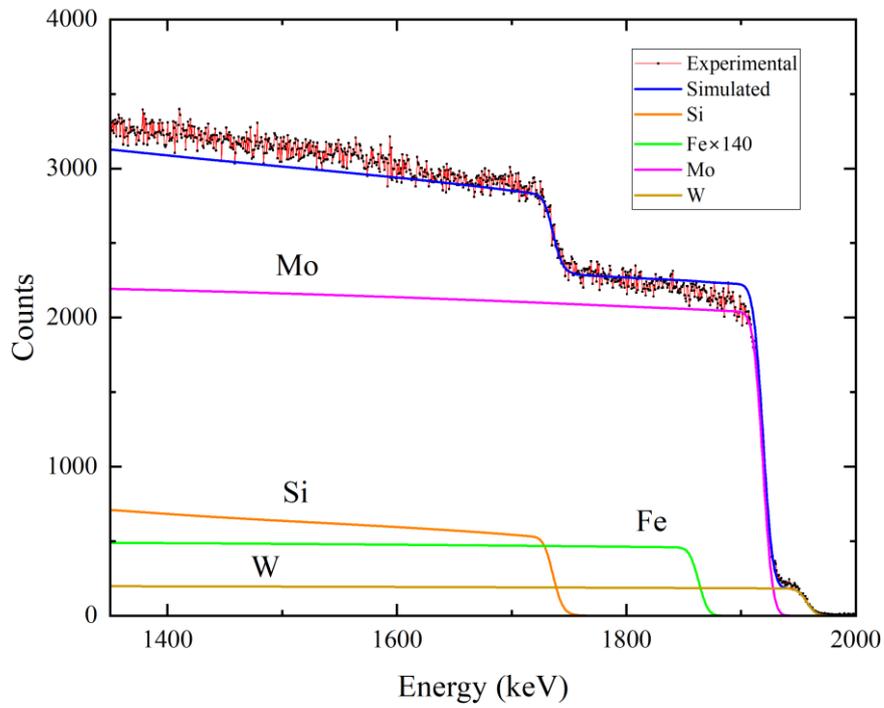

Figure 2. The EBS experimental and simulation spectra of MoSi$_2$ sample at 165 ° using 2 MeV proton impact. The counts of Fe is multiplied by 140 for visual aids.

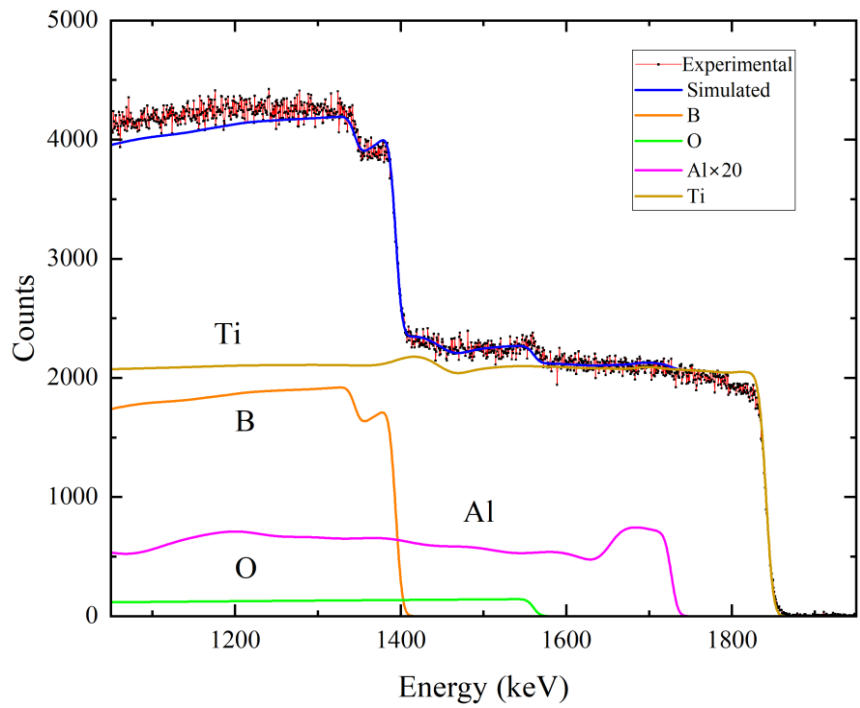

Figure 3. The EBS experimental and simulation spectra of TiB$_2$ sample at 165 ° using 2 MeV proton impact. The counts of Al is multiplied by 20 for visual aids.

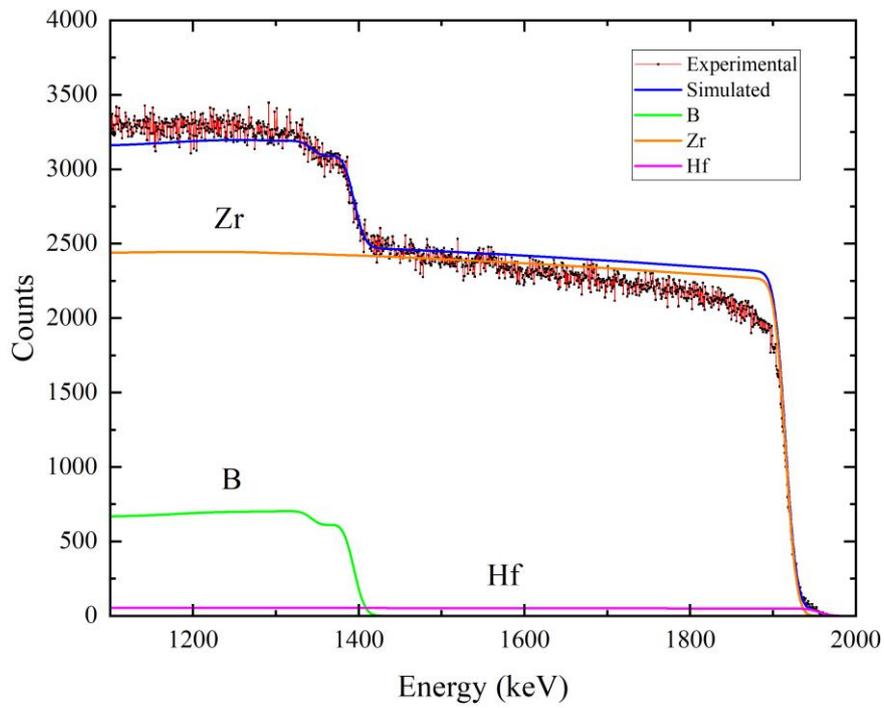

Figure 4. The EBS experimental and simulation spectra of ZrB$_2$ sample at 165 ° using 2 MeV proton impact.

The surface topography of the samples was observed by a scanning electron microscope (Helios G4 UC, Thermo Fisher Scientific Inc., USA), which is shown in Figure 5. The surface roughness was measured using an optical surface profilometers (Nano 3D Optical Surface Profilometers SuperView W1, Chotest Technology Inc., Shenzhen, China) with a scanning step length of 0.037 μm. Figures 6 and 7 show the surface profile and height distribution of three samples from the optical surface profilometers, and approximate Gaussian fittings to the height probability density functions. Table 1 shows the roughness parameters of the target surfaces.

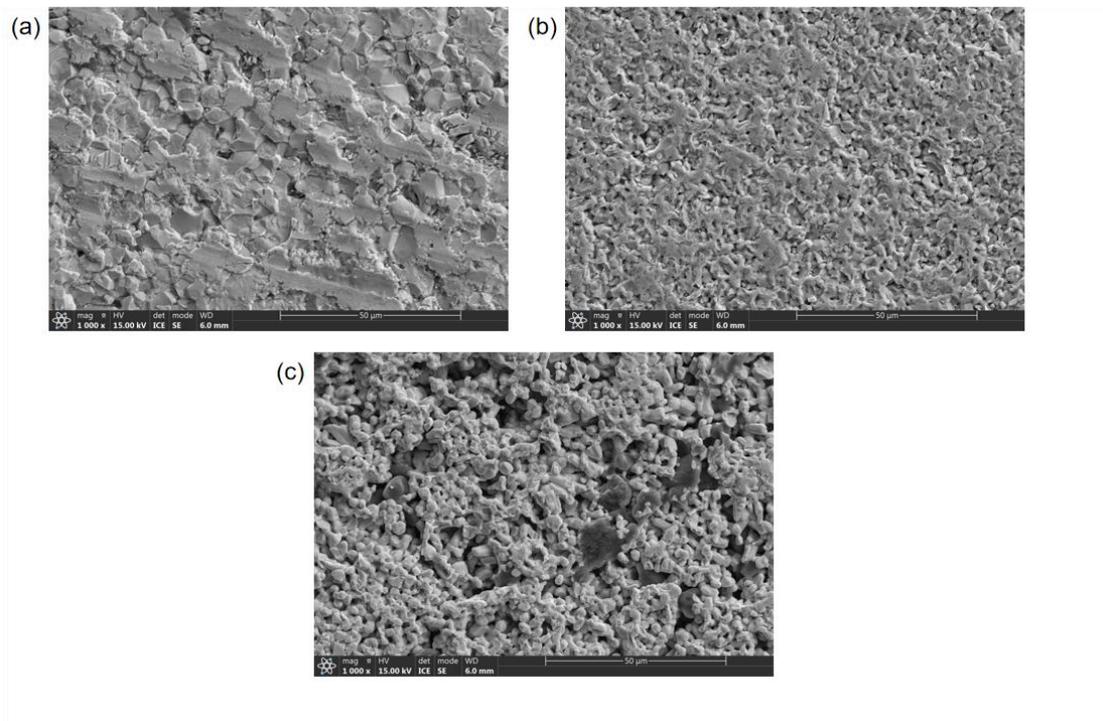

Figure 5. The SEM images of three sample surfaces. (a) MoSi$_2$, (b) TiB$_2$, (c) ZrB$_2$.

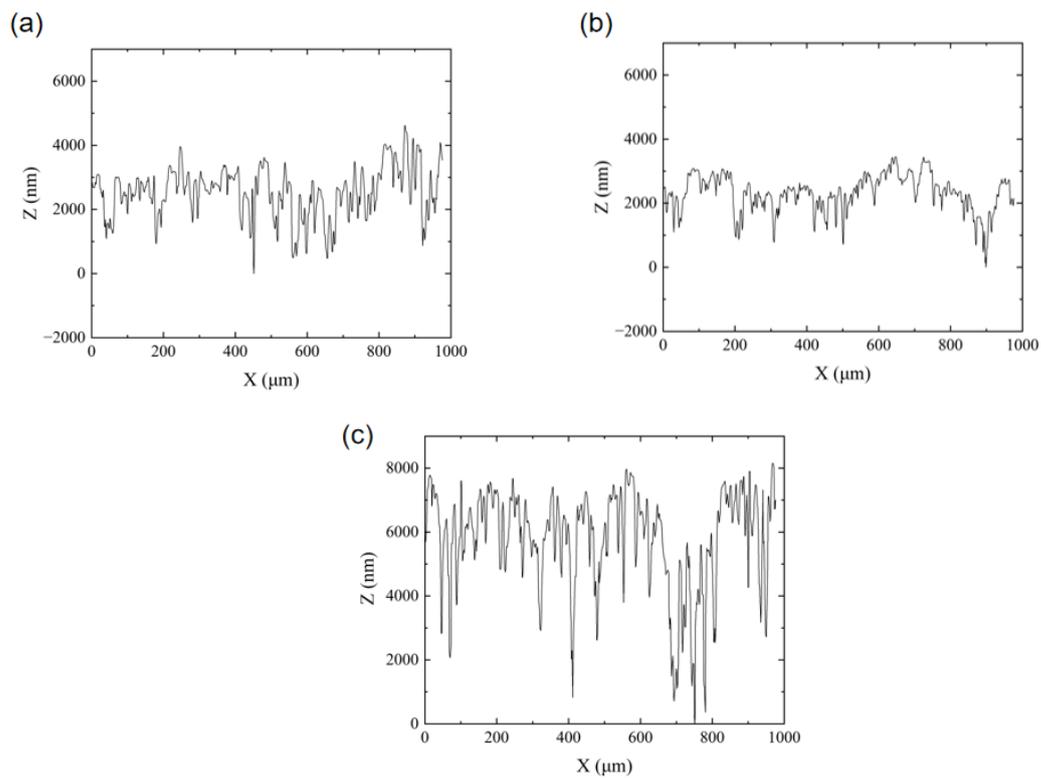

Figure 6. The surface profiles of three samples within the range of 1000 μm. (a) $MoSi_2$, (b) $TiB_2$, (c) $ZrB_2$.

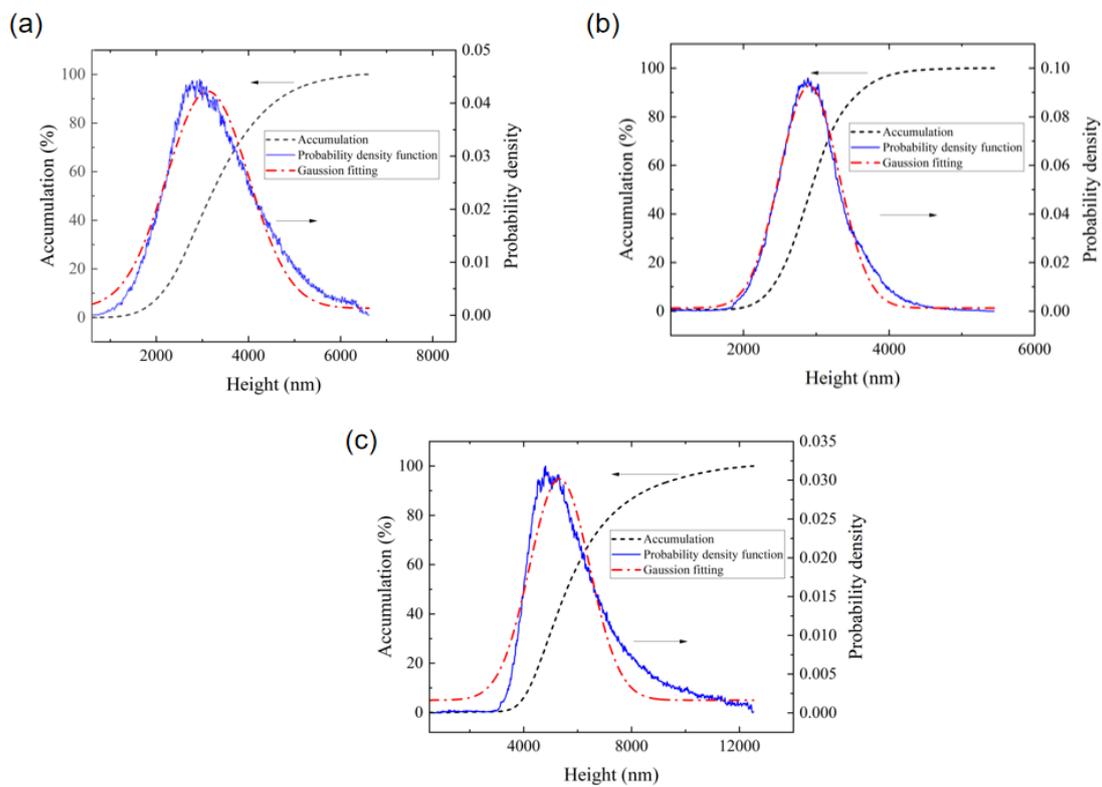

Figure 7. Surface height distributions with respect to the lowest valley of three sample surfaces: Cumulative distribution function, probability density function and approximately standard Gaussian fittings. (a) MoSi$_2$, (b) TiB$_2$, (c) ZrB$_2$.

Table1. Surface roughness parameters of the three samples. Sa is the arithmetical mean deviation of surface heights from the mean height, Sq is the root mean square deviation of surface heights, σ is the standard deviation of Gaussian fitting.

| Sample | Sa (nm) | Sq (nm) | σ (nm) |
| --- | --- | --- | --- |
| MoSi$_2$ | 796.7 | 1001.7 | 890.5 ±6.7 |
| TiB$_2$ | 376.9 | 504.5 | 406.2 ±1.6 |
| ZrB$_2$ | 1366.7 | 1776.2 | 1110.9 ±12.7 |

## 2.3 Monte Carlo simulations

The parallelized PENELOPE code [1], running on a computer cluster with 472 cores, 3.0 GHz operating frequency and 630 GB ROM, was used for the MC simulations, in which the experimental geometry was constructed and the elemental compositions of the samples, determined in the section 2.2, were employed. The half angle of X-ray impact detector subtended to the sample center was set as 5°. Moreover, when necessary, the surface roughness of the samples can be taken into account approximately by assuming a series of semi-spheres distributed on the sample's smooth surfaces [53,54]. The radii of the semi-spheres were determined by a Gaussian distribution, whose mean value and standard deviation, respectively, are equal to be the half of the full-width at one-tenth maximum (FWTM) and the standard deviation of the Gaussian distribution describing the surface roughness shown in Figure 7. A Fortran code was written to generate the geometry definition file used by the PENELOPE code. The radius of incident electron beam was set as the half of the radius of rough area. In addition, the default simulation parameters in the PENELOPE code, i.e., C1 and C2 for elastic scattering, WCC and WCR for cutoff energy losses and EABS for absorption energies, were used. The interaction forcing was also used to speed up the simulations.

The obtained simulation spectra were also convoluted with the Gaussian response function of SDD. The FWHM of the Gaussian response function can be simply expressed as follows:

$$\text{FWHM} = \sqrt{8WFE \ln 2 + \Delta E_{\text{elec}}^2} \ , \tag{1}$$

where W is the average energy for electron-hole creation (i.e., 3.62 eV), F is Fano factor (i.e., 0.12), E is the X-ray energy, and $\Delta E_{elec}$ is the electronic noise (i.e., 63.5 eV), which is determined based on the fact that the FWHM of the SDD is 135 eV at 5.89 keV.

## 2.4 Experimental data processing

The pile-up corrections for the measured X-ray spectra have been made. The correction algorithm was based on the Monte Carlo method proposed in Ref. [55] and described in detail in our previous paper [56]. The effects of pile-up correction are less than 1%. The escape rates of backscattered electrons (> 100 eV) from the top and side aperture of the Faraday cup were also estimated by using MC PENELOPE simulations with the experimental geometry. The escape rates for the three samples are, respectively, as follows: 2.66%-3.12% for $MoSi_2$, 2.24%-2.43% for $TiB_2$ and 2.57%-3.10% for $ZrB_2$ in the electron impact energies studied in this paper.

The experimental spectra were converted by the following formula for comparing with the MC simulation spectra:

$$N(E) = \frac{N_x(E)}{N_0 \Delta\Omega \varepsilon(E) \Delta E} , \qquad (2)$$

where $N_x(E)$ is the measured X-ray counts in a photon energy channel which have been corrected for the pile-up effect, $N_0$ is the incident electron number corrected by the escape rate, $\Delta\Omega$ is the solid angle subtended by the X-ray detector, $\varepsilon(E)$ is the intrinsic detection efficiency and $\Delta E$ is the photon energy channel width (31.01 eV/channel in this work). $\Delta\Omega\varepsilon(E)/4\pi$ is the absolute detection efficiency, $\Delta\Omega$ is determined as $2.75 \times 10^{-3}$ sr.

## 3. Results and discussion

The comparisons between the experiments and the MC PENELOPE simulations are shown in Figures 8-10. Only the comparisons for the X-ray energy region larger than 1 keV are concerned because the MC simulations for the X-ray energy region less than 1 keV cannot be considered as quantitative results [25]. The experimental uncertainties mainly come from the detector's efficiencies (~3%), sample compositions (~3%), incident charge measurements (~1%) and counting statistics (typically ~2-3%). The total experimental uncertainties in quadrature is ~5.3%. The uncertainties for the MC PENELOPE simulations are from the statistical uncertainties, typically less than 1% (1σ).

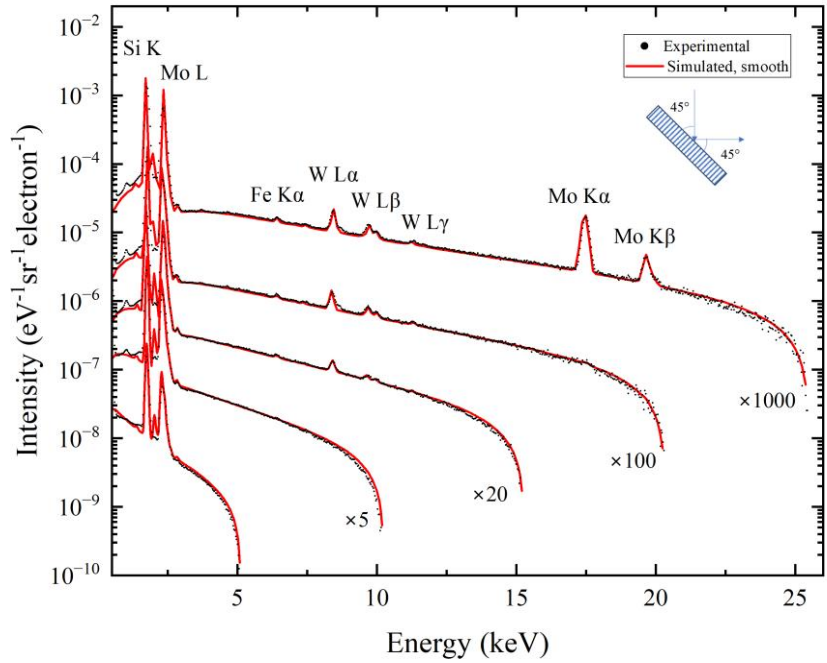

Figure 8. The comparisons between the experimental spectra and PENELOPE simulations for a thick $MoSi_2$ solid sample are presented. The dots represent the experimental spectra generated by the electron impact and the solid lines represent the simulation spectra. The incident electron energies are 5.10, 10.22, 15.24, 20.30, 25.44 keV with uncertainties of ~0.03 keV. Some spectra have been multiplied by factors as noted in the figure for visual aids.

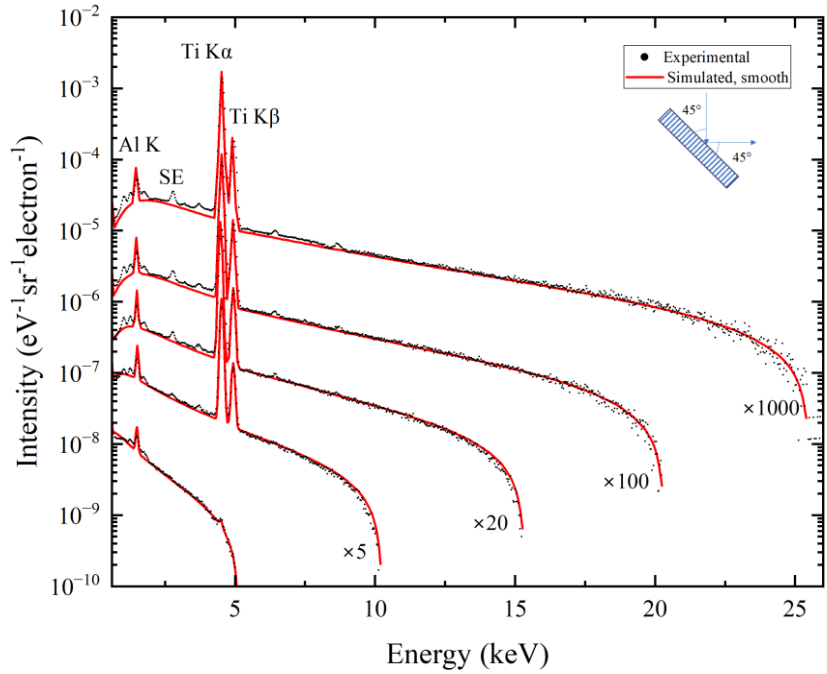

Figure 9. The same as in Figure 8, except for TiB2. The incident electron energies are 5.10, 10.22, 15.30, 20.30, 25.47 keV with uncertainties of ~0.03 keV. SE represents the single escape peak of Ti Kα.

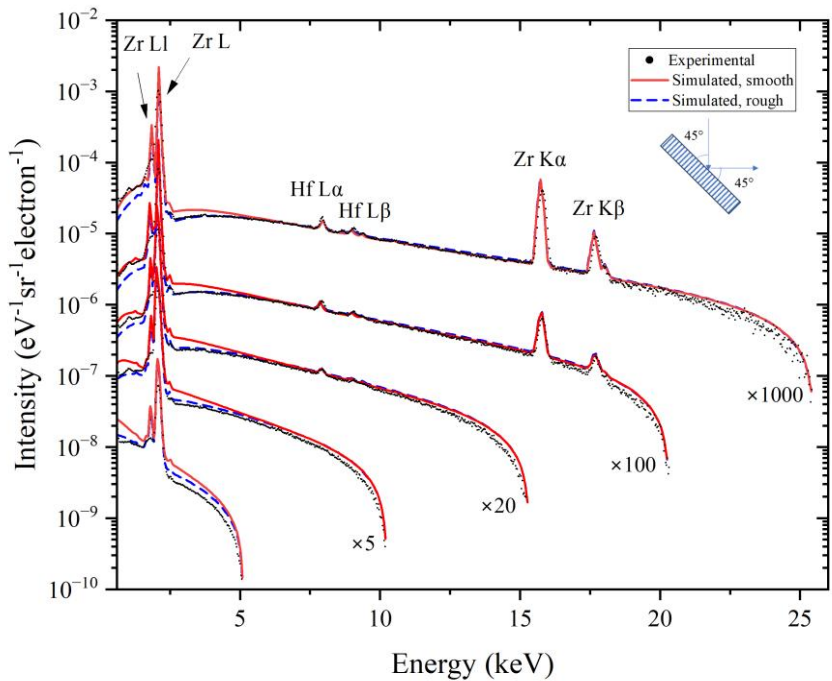

Figure 10. The same as in Figure 8, except for ZrB$_2$. The solid and dashed lines

represent the simulation spectra with the smooth and rough surfaces, respectively. The incident electron energies are 5.10, 10.22, 15.30, 20.30, 25.47 keV with uncertainties of ~0.03 keV.

In Figure 8, for the $MoSi_2$ sample, the experimental spectrum overall is ~5% lower than the MC simulation result for the incident electron energy of 5.10 keV. For other incident electron energies, as a whole, the experimental spectra and the MC simulation results are in good agreement, while at the higher X-ray energy region the experimental spectra are ~5% and ~10% lower than the MC simulation results for the incident electron energy of 15.24 keV and 10.22 keV, respectively. It can be said that within the experimental uncertainties the experimental spectra and MC simulation results are overall in agreement. The characteristic peaks from the impurities of W and Fe elements can be observed. The MC simulations are performed using smooth surfaces.

In Figure 9, for the $TiB_2$ sample, the experimental spectrum is ~5% higher than the MC simulation result at the low X-ray energy region for the incident electron energy of 25.24 keV. For other incident electron energies, as a whole, the experimental spectra and the MC simulation results are in good agreement, within ~3% differences. Overall, the experimental spectra and MC simulation results are in agreement within ~3-5%. The characteristic peak from the impurity of Al element, Ti K$\alpha$ single escape (SE) peak and some unidentified peaks can also be observed. The MC simulations are performed using smooth surfaces.

From Figure 5, the SEM images of three sample surfaces show that the surface roughness of $ZrB_2$ sample is very serious and the grains aggregate, while the surfaces of other two samples are smoother and can be well MC simulated using the smooth surfaces shown in Figures 8 and 9. For the $ZrB_2$ sample, in Figure 10, the experimental spectra and MC simulation results with smooth and rough surfaces are shown. The model of rough surface is described in Section 2.3. It can be seen the MC simulations using the smooth surface cannot well describe the experimental spectra, especially for lower incident electron energies (e.g., for 5.10 keV) and lower X-ray energy regions (e.g., less than ~5 keV). It is observed that the differences between the simulation results using the smooth and rough surfaces are located in the lower X-ray energy regions (i.e., less than ~5 keV). When using the model of rough surfaces, the agreement between the experiments and the MC simulations are improved to some extent. For the incident electron energies of 5.10 keV and 10.22 keV, the differences between the experimental spectra and the simulation spectra are ~20% and ~10%, respectively. For other incident electron energies, the differences are ~5-10%, mainly in the higher X-ray energy region, while in the other X-ray energy region the agreement is good within ~5% differences. In general, for the higher incident energies (> 15 keV), the experiments and the MC simulations can be in agreement with ~5% differences, except in the higher X-ray energy region. It should be kept in mind that the model of rough surface in this work is very approximate, and more realistic roughness model [57] should be utilized for more accurate simulations. Due to the serious surface roughness of $ZrB_2$ target and the approximation of the roughness model, $ZrB_2$ target was not used to evaluate the validity of the additivity approximation.

## 4. Conclusions

Experiments have been performed to measure the bremsstrahlung spectra produced by ~5-25 keV electrons incident on $MoSi_2$, $TiB_2$ and $ZrB_2$ solid conductive compound samples in the way of absolute measurements. These samples were characterized by various analysis methods (i.e., EBS, PIXE, WD-EPMA, SEM and Surface Profilometer). The experimental spectra was corrected for the signal pileup effect and incident electron escape. The uncertainties for the measured bremsstrahlung spectra are ~5.3%. The comparisons between the experiments and MC PENELOPE simulations are made. It is shown that the experimental bremsstrahlung spectra and the MC PENELOPE simulations overall are in agreement within the experimental uncertainties for $MoSi_2$ and $TiB_2$ samples at all incident electron energies. These results show that within the experimental uncertainties the MC PENELOPE can simulate well the experimental spectra for compounds and mixtures and the additivity approximation in Monte Carlo simulations is feasible, at least, for the studied cases in this work. And the significant differences between experiments and Monte Carlo simulations near the Duane-Hunt limit for solid insulating compound samples in previous studies [28-30] do not appear in this studies. In the future, more experimental investigations should be performed for various compounds and mixtures.

**Acknowledgments**

The financial support from the National Natural Science Foundation of China

under Grant No.12175158 is acknowledged. The staffs at Institute of Nuclear Science and Technology, Sichuan University, F. Bai, P. Wang, Y.X. Fan and G. Liu, are appreciated for operating the 3 MV tandetron accelerator. The authors would like to thank Profs. Z.H. Li and K. Zhang for their helps in this study.## References

[1] L. Li, Z. An, J. Zhu, W. Lin, S. Williams, Nuclear Instruments and Methods in Physics Research Section B: Beam Interactions with Materials and Atoms 506 (2021) 15.
[2] P.A. García-Higueras, S. García-Pareja, F. Salvat, A.M. Lallena, Radiation Physics and Chemistry 208 (2023) 110949.
[3] D. Gonzales, S. Requena, S. Williams, Applied Radiation and Isotopes 70 (2012) 301.
[4] S. Requena, S. Williams, C.A. Quarles, Nuclear Instruments and Methods in Physics Research Section B: Beam Interactions with Materials and Atoms 268 (2010) 3561.
[5] S. Czarnecki, A. Short, S. Williams, Nuclear Instruments and Methods in Physics Research Section B: Beam Interactions with Materials and Atoms 378 (2016) 54.
[6] B. Singh, S. Prajapati, S. Kumar, B.K. Singh, X. Llovet, R. Shanker, Radiation Physics and Chemistry 150 (2018) 82.
[7] B. Singh, S. Prajapati, B.K. Singh, R. Shanker, Applied Radiation and Isotopes 148 (2019) 126.
[8] A. Singh, A.S. Dhaliwal, Applied Radiation and Isotopes 115 (2016) 190.
[9] E.V. Gnatchenko, A.N. Nechay, A.A. Tkachenko, Physical Review A 80 (2009) 022707.
[10] S. Prajapati, B. Singh, B.K. Singh, and R. Shanker, Radiation Physics and Chemistry 153 (2018) 92.
[11] S. Prajapati, B. Singh, B.K. Singh, and R. Shanker, The European Physical Journal D 73 (2019) 155.
[12] S. Prajapati, B. Singh, S. Kumar, B.K. Singh, and R. Shanker, Journal of Physics. B, Atomic, Molecular, and Optical Physics 52 (2019) 145201.
[13] R. Shanker and R. Hippler, Zeitschrift für Physik D-Atoms Molecules and Clusters 42 (1997) 161.
[14] C. Quarles and L. Estep, Phys Lett A 110 (1985) 387.
[15] C. Quarles and L. Estep, Phys Lett A 114 (1986) 9.
[16] L. Tian, J. Zhu, M. Liu, Z. An, Nuclear Instruments and Methods in Physics Research Section B: Beam Interactions with Materials and Atoms 267 (2009) 3495.
[17] L. Li, Z. An, J. Zhu, W. Tan, Q. Sun, and M. Liu, Physical Review. A 99 (2019) 052701.
[18] L. Li, Z. An, J. Zhu, M. Liu, Nuclear Instruments and Methods in Physics Research Section B: Beam Interactions with Materials and Atoms 445 (2019) 13.
18